# SHORT-TIME EVOLUTION OF ALKANE-IN-WATER NANO-EMULSIONS

German Urbina-Villalba*, Kareem Rahn-Chique

Instituto Venezolano de Investigaciones Científicas (IVIC), Centro de Estudios Interdisciplinarios de la Física (CEIF), Laboratorio de Fisicoquímica de Coloides, Carretera Panamericana Km. 11, Altos de Pipe, Edo. Miranda, Apdo. 20632, Caracas, Venezuela. e-mail: guv@ivic.gob.ve

**Abstract**　The stability of alkane-in-water nanoemulsions during the sub-stationary regime is studied by means of Emulsion Stability Simulations (ESS). The effects of Ostwald ripening, flocculation, coalescence, gravity, and hydration forces are considered. According to these calculations flocculation and coalescence are predominant during the first few seconds after the preparation of the emulsion. This favors the generation of a right-skewed Drop Size Distribution (DSD). As the system evolves, the drops grow larger and more repulsive causing a slow down of the flocculation process. In the case of dodecane ($C_{12}$) and hexadecane ($C_{16}$) the referred phenomena, reinforce the ripening trend to subvert the initial DSD variation, producing a meta-stable distribution which is preserved during several minutes. After this time, Ostwald ripening dominates: the skirt of the distribution changes progressively from right-skewed to left-skewed. Consistent with these changes, the cube average radius ($R_a^3$) of the emulsion increases rapidly at first, but progressively diminishes generating a concave-downward curve that stabilizes asymptotically. In the case of dodecane and hexadecane the complete dissolution of the drops promoted by ripening is prevented at all times due to coalescence. In the case of octane ($C_8$) a substantial amount of drops is lost by dissolution, forbidding the attainment of a stable DSD. In all cases the molecular exchange *only favors a decrease of the average radius as a function of time*. It is the elimination of drops either by dissolution or coalescence which causes an increase of the average radius of the emulsion.

**Keywords**　Nano, Emulsion, Dodecane, Octane, Hexadecane, Ripening, Stability.

## 1. INTRODUCTION

The exchange of molecules between drops of different sizes is a well known phenomenon, whose main features are justified on the basis of the Lifshitz-Slyozov-Wagner (LSW) theory [Lifshitz, 1961; Wagner, 1961]. According to LSW, the Ostwald ripening rate ($V_{OR}$) has a constant value in the asymptotic limit, given by:

$$V_{OR} = dR_c^3/dt = 4\alpha\, D_m\, C(\infty)/9 \qquad (1)$$

Where $R_c$, $D_m$, $C(\infty)$ and $\alpha$ stand for the critical radius of the dispersion, the diffusion constant of the oil molecules, their bulk solubility in the presence of a planar Oil/Water (O/W) interface, and the capillary length, defined as:

$$\alpha = 2\gamma\, V_m / \tilde{R}\, T \qquad (2)$$

Here $\gamma$ is the O/W interfacial tension, $V_m$ the molar volume of the oil, $\tilde{R}$ the universal gas constant, and $T$ the absolute temperature. Finsy [2004] demonstrated that the critical radius of the dispersion is equal to its number average radius ($R_a$):

$$R_c = R_a = \frac{1}{N_T} \sum_k R_k \qquad (3)$$

LSW envisages the long-time development of a characteristic left-skewed drop-size distribution with a cut-off radius of $1.5 R_c$. This distribution results from exchange of oil molecules between drops. Such exchange occurs by diffusion through the aqueous phase. Drops with radii smaller than the critical radius decrease their size, and those with radii larger than $R_c$ increase their size. Since the critical radius changes as a function of time, theory predicts that the drops are constantly dissolving and growing, favouring the development of a self-similar DSD at long times.

The LSW distribution has only been observed in systems with surfactants [Schmitt, 2004] where the processes of flocculation and coalescence are substantially decelerated. This is consistent with two of the main as-

German Urbina-Villalba, Kareem Rahn-Chique

sumptions of this theory: 1) the particles of the system are fixed in the space, and 2) the system is infinitely dilute. These suppositions avoid the consideration of a direct interaction between the drops and their possible exchange of oil due to coalescence.

Alkane-in-water emulsions are convenient systems to test the validity of the LSW theory. In the absence of surfactants, these emulsions should expose the essential features of Ostwald ripening related to the oil drops alone. Unfortunately neat oil drops exhibit a significant electrostatic surface potential due to the preferential adsorption of hydroxyl ions to the oil/water interface [Stachurski, 1996; Beattie, 2004]. However, even in this case, the electrostatic potential should prevent the flocculation of the drops reinforcing the theoretical predictions of LSW. Therefore it is surprising that emulsions of cyclohexane, n-hexane, n-octane, n-decane, n-tetradecane, and n-hexadecane, exhibit ripening rates ($V_{OR}$) which are 32, 97, 679, 8, 3, and 17 times larger than the theoretical prediction [Sakai, 2002].

Sakai *et al.* [2001] demonstrated the occurrence of coalescence in benzene in water emulsions. Using freeze-fracture electron microscopy, the authors observed small drops with diameters at 30-100 nm immediately after sonication, and aggregates of medium size (200-500 nm) composed of small droplets. An hour later, the small drops had coalesced into larger drops. Medium size drops also aggregate between themselves and subsequently coalesce (> 1000 nm). A possible explanation of these results was proposed by Sakai *et al.* [2003] based on the mean free path between the drops. According to the authors the drops of each generation flocculate with similar drops and subsequently coalesce inside the aggregates producing larger drops. The new drops (next generation) also aggregate among them and then coalesce. As a result, the radii of the drops vary geometrically with a common factor of $(3\phi)^{-1/3}$, where $\phi$ is the volume fraction of oil.

The progressive decrease of the number of particles per unit volume at constant volume fraction $\phi$ which occurs as a consequence of the coalescence of the drops is not obvious and may have interesting consequences. If the coalescence between the drops occurs step by step with the progressive addition of single particles to the cluster –as was envisioned by Smoluchowski [1917]–, a drop of size k will have a radius of $R_k = \sqrt[3]{k} R_0$. However, if the drops of the same size aggregate first and then coalesce to form a bigger drop, the radius of the drop changes according to [Sakai, 2003]:

$$R_{k-1} = (1/\phi)^{(k-1)/3} R_0 \qquad (4)$$

Hence, the radii of the drops changes in a geometric series, and as a result the number of particles per unit volume decreases substantially with the increase of the particle size.

The model of Sakai *et al.* [2003] has the disadvantage that it assumes that only particles of the same "generation" (size) aggregate to form a bigger drop. However, if it is considered that the diffusion constant of the drops decreases with size, and the mean free path increases as the number of particles decreases, then it seems reasonable that the particles of smaller size aggregate and coalesce substantially faster than larger drops. As a consequence only drops of each generation aggregate (and coalesce).

In previous communications our group examined the influence of flocculation and coalescence on the short-time evolution of the average radius of a dodecane-in-water nanomulsion using both experimental measurements and simulations [Urbina-Villalba, 2009a, 2009b, 2009c; Urbina-Villalba, 2012]. It was found that:

a. A repulsive potential between the drops causes a concave downward dependence of the cube average radius of the emulsion as a function of time. The slope of $dR_a^3/dt$ progressively decreases, slowly approaching the LSW limit.
b. For very short times (6 s < t < 20 s) and in the case of a dodecane-in-water emulsion, ESS predicts a rate ($dR_a^3/dt$) of $(1.0 \pm 0.5) \times 10^{-24}$ m$^3$/s ($r^2$ = 0.9462). This value agrees with the experimental measurement at a concentration of 0 mM NaCl [Urbina-Villalba *et al.*, 2009a]. Such variation is essentially caused by flocculation and coalescence.
c. For t < 100 s $dR_a^3/dt$ increases appreciably with the augment of the salt concentration. This behavior is consistent with the theory of Derjaguin, Landau, Verwey and Overbeek [1946] evidencing the influence of flocculation in the destabilization process.
d. At long times (t >> 200 s) $dR_a^3/dt$ attains values of the order of $10^{-26}$ m$^3$/s. This value is similar to the one reported by Sakai: 3.09 x $10^{-26}$ m$^3$/s for a dodecane-in-water emulsion, but three times higher than the LSW prediction.

e. According to ESS, the shape of the DSD of an oil-in-water emulsion drastically changes during the first 20 seconds after the preparation of the dispersion. Starting from a Gaussian distribution, it rapidly develops a skirt towards the right as a consequence of flocculation and coalescence. This variation is consistent with the findings of Sakai *et al.* [2002] for t = 4 minutes. At longer times, the simulations predict a stabilization of the distribution for several minutes, with the progressive development of a left-skewed skirt.

f. According to ESS, the increase of the average radius of the emulsion only occurs when the total number of particles of the system diminishes either due to their complete dissolution or to the coalescence of drops. The exchange of molecules due to Ostwald ripening *only decreases the average radius of the emulsion.* In the case of dodecane, the rate of flocculation and coalescence is faster than the one of ripening at least during several minutes, preventing the complete dissolution of the drops by means of molecular exchange.

g. When a high salt concentration is used to screen the electrostatic repulsion between the drops, the simulations predict a linear dependence of $(3.031 \pm 0.002) \times 10^{-22}$ m$^3$/s ($r^2 = 0.9996$). The referred slope is one order of magnitude higher than the one exhibited by a dodecane-in-water system at 600 mM ($2.9 \times 10^{-23}$ m$^3$/s). Hence there is an unknown repulsive potential which slows down the process of destabilization, but it is only effective at high ionic strengths.

In order to understand the differences between the values of $dR_a^3/dt$ obtained by Sakai *et al.* [2002] and those predicted by LSW, knowledge of transient period of time between the making of the emulsion and the attainment of the stationary limit of Ostwald ripening is required. In this regard, this article presents the predictions of ESS for the short-time evolution (t < 5 minutes) of three alkane-in-water nanoemulsions in the absence of surfactants. First, the role of flocculation and coalescence on the variation of $dR_a^3/dt$ is established. Second, the influence of gravity and hydration forces on the slope of $R_a^3$ vs. t is studied. Third, a comparison between the DSDs predicted by ESS for t = 4 minutes, and the ones measured by Sakai *et al.* [2002] for the referred systems is presented.

It should be remarked that the initial DSD resulting after the preparation of Sakai *et al*.'s emulsion is unknown.

Hence, the simulations start from a Gaussian distribution of drops to avoid any predisposition of the initial condition to generate a right-skewed or left-skewed DSD after some time. The average radius of the distribution and its width are also unknown, and were arbitrarily set to 30 nm and 1.5 nm, respectively. Due to all these limitations, our calculations are only expected to reproduce some general features of the drop size distribution at t = 4 minutes. However, the simulations are able to illustrate the underlying mechanism of evolution of the system, and make reasonable predictions about its long time behavior.

The paper is organized as follows. First a brief overview of ESS is shown. Then, the computational details of the calculations are specified. Subsequently, the results are presented along with the discussion and some concluding remarks.

## 2. EMULSION STABILITY SIMULATIONS

A detailed description of the algorithm of Emulsion Stability Simulations (ESS) can be found in references [Urbina-Villalba *et al.*, 2000, 2003, 2004, 2005, 2006, 2009a, 2009b, 2009c, 2010, 2012]. Here, only the essential aspects of the simulations are reviewed in order to illustrate the significance of the calculations.

ESS start from a cubic box that contains N drops randomly distributed. The particles move with an equation of motion similar to that of Brownian dynamics simulations:

$$\vec{r}_i(t + \Delta t) - \vec{r}_i(t) = \left\{ \left( \sum_{\substack{j=1 \\ j \neq i}}^{N} \vec{F}_{ji} + \vec{F}_{ext} \right) D_{eff,i}(d, \phi) \middle/ k_B T \right\} \Delta t + \sqrt{2\, D_{eff,i}(d, \phi) \Delta t} \, [\vec{G}\text{auss}] \quad (5)$$

The displacement of particle "i" during the time step $\Delta t$: $\vec{r}_i(t + \Delta t) - \vec{r}_i(t)$, is the result of two contributions: 1) deterministic inter-particle forces $\sum_{i=1; j \neq i}^{N} \vec{F}_{ji}$ and external forces $\vec{F}_{ext}$, acting on particle i; and 2) the random deviates produced by the interaction of the solvent with the moving particles. The stochastic vector Gauss stands for a set of random numbers, which have a Gaussian distribution with zero mean and unit variance. The characteristic mean square displacement of the Brownian movement is obtained multiplying each random deviate by $\sqrt{2 D_{eff}(d, \phi) \Delta t}$, where $D_{eff,i}(d, \phi)$ is the effective diffusion constant of drop i. In the case of non-deformable drops it is equal to:

$$D_{eff\ i}(d, \phi) = D_0\ f_{corr\ i} = (k_B T / 6 \pi \eta R_i) f_{corr\ i} \quad (6)$$

Where $\eta$ is the shear viscosity of the solvent, $R_i$ is the radius of particle i, and $k_B$ the Boltzmann constant. The correction factor $f_{corr\ i}$ depends on the total volume fraction of oil in the system ($\phi$), and the distance from its closest neighbour (d). At every time step of the simulation the program locates the position of the nearest neighbour of each particle *i*. If that particle gets within a distance of one radius from particle i: $d = r_{ij} - R_i - R_j = R_i (r_{ij} = |\vec{r}_i(t) - \vec{r}_j(t)|)$, the formula of Honig *et al* [1971] is used to correct the diffusion constant of i:

$$f_{corr} = (6 u^2 + 4 u)/(6 u^2 + 13 u + 2) \quad (7)$$

Where: $u = (r_{ij} - R_i - R_j) / R_R$, and $R_R = 1/2 (R_i + R_j)$. Otherwise the volume fraction of particles around i is used to evaluate an empirical expression of the diffusion [Beenakker, 1982; van Mengen, 1987]:

$$f_{corr} = 1.0 - 1.734 \phi + 0.91 \phi^2 \quad (8)$$

At the beginning of the simulation the program creates a Gaussian distribution of particle sizes. These particles are distributed randomly inside the simulation box. Next the program calculates the interfacial parameters of the drops (like surface charges and interfacial tension in this case). Then it assigns an effective diffusion constant to each particle. Following it computes the forces between the particles and moves them. The exchange of molecules due to Ostwald ripening proceeds at this point (see below). Finally, the program checks for the coalescence of drops. Coalescence occurs whenever the distance of approach between the drops is smaller than the sum of their radii. In this case, a new drop is created at the centre of mass of the coalescing drops. The new radius results from the conservation of volume: $R_{new} = \sqrt[3]{R_i^3 + R_j^3}$.

In order to mimic the Ostwald ripening process, the algorithm of De Smet *et al*. [De Smet, 1997] was implemented in the former code of ESS. The fundamental equation of this method is derived from Fick's law and Kelvin's equation assuming $\alpha \ll R_i$:

$$m_i(t + \Delta t) = m_i(t) + M(t) P_i(t) \quad (9)$$

Here, $m_i$ stands for the number of molecules of oil in particle i. $P_i(t)$ represents the growth law:

$$P_i(t) = R_i(t) / R_a(t) - 1 \quad (10)$$

According to Eqs. (9) and (10), particles with radii smaller than the average radius ($R_i < R_a$) dissolve, while particles with radii larger than the average radius ($R_i > R_a$) grow. Particles with the same radius as the average radius $R_i = R_a$, preserve their size. The average radius evolves during the simulation as the particles coalesce or dissolve.

According to Eq. (9), the number of molecules of oil exchanged by the particle i at time t, is equal to the product $M(t)P_i(t)$. $M(t)$ has a constant value during the simulation except when the smallest drop of the system (i = S) contains fewer molecules than the product $M(t) P_S(t)$. In this case, the value of $M(t)$ changes, so that only a fraction of the molecules of the smallest particle is exchanged:

$$M(t) = M = 4 \pi D_m C(\infty) \alpha \Delta t \quad m_S \geq M P_s(t) \quad (11)$$

$$M(t) = m_s / M \quad m_s < M P_s(t) \quad (12)$$

When the condition $m_s < M P_S(t)$ occurs drop i = S is eliminated after the exchange of oil molecules finishes, and its former molecules are redistributed among the surviving drops using Eq. (9). This avoids the recursive use of Eq. (12) for infinitesimal amounts of oil.

In a typical ESS, the number of particles progressively decreases as a result of the coalescence of drops. However, the process of Ostwald ripening requires maintaining a minimum number of drops during the simulation. This is especially relevant if the lapse of time to be simulated is long. In order to solve this problem a special procedure for the "regeneration" of the drop size distribution was implemented. The simulations start with a number of drops equal to $N(t=0) = N_0 = 500$. As the drops coalesce and dissolve, the number of drops decreases until $N(t=t') = N_0/4$. At this point, the entire simulation box, originally centred at $(x,y,z) = (0,0,0)$, is translated to the negative quadrant of the coordinate axis. Following it is replicated three times using periodic boundary translations. The new simulation box comprises the former box and its three replicas. Hence, it preserves the volume fraction of oil ($\phi = \phi (t=0)$), the relative coordinates of the particles, and the DSD at

t=t'. Moreover, it restores the number of particles to its initial value, maintaining the statistical significance of the ripening procedure. The algorithm is convenient -even in the absence of ripening- to reproduce the evolution of the system at long times.

In the present calculations, the following potentials were employed:

a) van der Waals [Hamaker, 1937]:

$$V_A = V_{vdW} = -A_H/12 \left( y/(x^2 + xy + x) + y/(x^2 + xy + x + y) \right.$$

$$\left. + 2\ln\left[(x^2 + xy + x)/(x^2 + xy + x + y)\right] \right) \quad (13)$$

Here: $x = h/2R_i$, $y = R_i/R_j$, $h = r_{ij} - R_i - R_j$, and $A_H$ is the Hamaker constant.

b) Electrostatic [Danov, 1993]

$$V_E = (64\pi C_{el} k_B T/\kappa) \tanh(e\Psi_{0i}/4k_B T) \tanh(e\Psi_{0j}/4k_B T) \times$$

$$\exp(-\kappa h) \left[ 2 R_i R_j / \kappa(R_i + R_i) \right] \quad (14)$$

Where: $C_{el}$ is the concentration of electrolyte, $\kappa^{-1}$ is the Debye length, and e the unit of electrostatic charge.

c) Hydration (spherical drops) [Ivanov, 1999]:

$$V_{hyd} = \pi R_i \lambda_0 f_0 \exp(-h/\lambda_0) \quad (15)$$

With $\lambda_0 = 0.6$ nm, and $f_0 = 3$ mJ/m².

The forces result from the differentiation of these potentials with respect to the distance between the drops. Some of the calculations include the buoyancy force:

$$F_{ext} = F_b = \frac{4}{3}\pi R_i^3 \Delta\rho\, g \quad (16)$$

Here $\Delta\rho$ stands for the density difference between dodecane and water, and g is the gravity force. 3-D periodic boundary conditions are used in all cases. In this regard it should be noticed that the simulation box is of the order of the size of the particles (L = 209.6 R, R = 30 nm). Hence, the particles cannot accumulate at the top of the box as it occurs in a macroscopic container. Instead, the particles that "leave" the simulation box through its upper face are reinserted through its lower face and vice-versa.

This mimics the situation that occurs in a small volume of the experimental system located at the middle of the container: Some particles escape the slice through its upper facet due to their Brownian movement and the buoyancy force, but at the same time, some other particles enter the slice through its bottom. This does not happen at the top of a macroscopic container where the particles accumulate, or at its bottom, where the volume is depleted from the larger particles.

**3. COMPUTATIONAL DETAILS**

All the simulations start from a Gaussian distribution of particle sizes centered at $R_a$ = 30 nm (standard deviation of 1.5 nm). The average particle size corresponds to the lowest value observed by Sakai *et al.* [2002] in their experiments. It is implicitly assumed that this size can always be reached by the emulsification procedure, and it is the evolution of the emulsion that produces the distinct characteristics of the drop size distribution observed at t = 4 minutes.

The initial number of particles in the simulation is $N_T = N_0 = 500$. The size of the initial simulation box was adjusted so that the volume fraction of oil was equal to $\phi = 2.29 \times 10^{-4}$ in all cases. The routine of Ostwald ripening was activated in all cases. Hence, the number of particles never decreases below 125. As previously explained, when $N(t') = 125 = N_0/4$, the DSD is reconstructed and the number of particles incremented to N = 500.

The time step of the calculations was set very short ($\Delta t = 1.37 \times 10^{-7}$ s) in order to sample the interaction potentials appropriately. Due to the long tail of the electrostatic potential a long cut off of 750 nm was used. Even with these approximations, the simulations lasted for more than 2 years in a Dell Precision workstation with 8 processors.

Beattie and Djerdjev [Beattie, 2004] demonstrated that hydroxide ions are the specific charge-generating species at the hexadecane oil/water interface. In the absence of surfactants, the pH of the alkane-in-water solutions depends on the total interfacial area of the emulsion. The addition of NaOH or HCl to reproduce a specific pH, changes the surface potential ($\Psi_0$) of the drops considerably, allowing variations that go from -40 mV to -120 mV for pHs between 5 and 8. The paper of Sakai *et al.* [2002] does not specify the electrostatic characteristics of the suspended drops, or the pH of their emulsions. In previous dodecane-in-water

German Urbina-Villalba, Kareem Rahn-Chique

emulsions of similar volume fraction ($\phi \sim 10^{-4}$) synthesized by our group [Urbina-Villalba, 2009a], a pH ~ 6, was observed. Hence, we set the ionic strength of the solutions in these calculations to $10^{-6}$ M. Using this ionic strength (I), a surface charge density of $\sigma = -0.3$ mCoul/m$^2$ produces a surface potential of -11.6 mV. This value was calculated by the program using the analytical expression of Sader [1995; 1997] with $R_i$ = 30 nm:

$$\sigma_T e / \kappa \varepsilon \varepsilon_0 k_B T = \Phi_p + \Phi_p / \kappa R_i - \kappa R_i (2 \sinh(\Phi_p/2) - \Phi_p)^2 / \overline{Q}$$

$$\overline{Q} = 4 \tanh(\Phi_p/4) - \Phi_p - \kappa R_i [2 \sinh(\Phi_p/4) - \Phi_p] \quad (17)$$

Where: $\Phi_p = \Psi_0 e/k_B T$ is the reduced electrostatic potential of the particle at its surface, $\varepsilon_0$ is the permittivity of vacuum and $\varepsilon$ the dielectric constant of water.

For the present calculations only spherical non-deformable drops were used. The following systems were studied:

1) $C_{12}$ (E): Dodecane-in-water droplets with a surface potential of $\Psi_0$ = -11.6 mV.
2) $C_{12}$ (E,G): Similar to $C_{12}$(E) with the effect of the gravity included.
3) $C_{12}$ (E,H): Similar to $C_{12}$(E) with the effect of the hydration forces included.
4) $C_8$ (E): Octane-in-water droplets with a surface potential of $\Psi_0$ = -11.6 mV.
5) $C_8$ (E,G): Similar to $C_8$(E) with the effect of the gravity included.
6) $C_8$ (E,H): Similar to $C_8$(E) with the effect of the hydration forces included.
7) $C_{16}$ (E): Hexadecane-in-water droplets with a surface potential of $\Psi_0$ = -11.6 mV.
8) $C_{16}$ (E,G): Similar to $C_{16}$(E)$_1$ with the effect of the gravity included.
9) $C_{16}$ (E,H): Similar to $C_{16}$(E)$_1$ with the effect of the hydration forces included.

The parameters of the simulations are shown in Table 1. Notice that the same electrostatic charge was used regardless of the type of oil. Since the Hamaker constants of these oils are similar, the use of the same electrostatic and hydration potentials, which allows identifying differences in the ripening rates. Those differences are due to the solubility of the oils which roughly increases in two orders of magnitude as the number of carbon atoms of the oil is augmented in four units. However, it must be remembered that the surface charge of the oil changes with the type of oil. Those differences are disregarded in the present simulations.

Table 1: Parameters of the Simulations

| Property | Octane | Dodecane | Hexadecane |
|---|---|---|---|
| $A_H$ (J) | 4.10 x 10$^{-21}$ | 5.02 x 10$^{-21}$ | 5.40 x 10$^{-21}$ |
| C ($\infty$) cm$^3$/ cm$^3$ | 9.42 x 10$^{-7}$ | 5.31 x 10$^{-9}$ | 2.72 x 10$^{-11}$ |
| $D_m$ (m$^2$/s) | 7.80 x 10$^{-10}$ | 5.40 x 10$^{-10}$ | 4.60 x 10$^{-10}$ |
| $V_m$ (m$^3$/mol) | 1.63 x 10$^{-4}$ | 2.27 x 10$^{-4}$ | 2.92 x 10$^{-4}$ |
| $\gamma$ (mN/m) | 51.68 | 52.78 | 53.77 |
| $\sigma_1$ (mCoul/m$^2$) | -0.30 | -0.30 | -0.30 |
| $\rho$ (g/cm$^3$) | 0.703 | 0.749 | 0.775 |
| $f_0$ (mJ/m$^2$) | 3 | 3 | 3 |
| $\lambda$ (nm) | 0.6 | 0.6 | 0.6 |

Figure 1 shows the potentials of interaction between two drops of dodecane suspended in water. Due to the low ionic strength, the electrostatic potential shows a long decaying length that extends well beyond 100 nm ($V_E$ = 1 $k_B$T at 375 nm). This potential requires a long cut-off length. Instead, the van der Waals potential and the hydration potential are only appreciable below 4 nm. Therefore, the time step should be sufficiently short to sample those potentials ap-propriately. The barrier of the electrostatic potential is lower ($4\,k_B T$) than the one of the hydration potential ($20\,k_B T$). Curiously there is some sort of secondary minimum ($\Delta V \sim 0.19\,k_B T$) at h = 3.6 nm (where h is the shortest distance between the particles' surfaces. As expected, the potentials corresponding to octane and hexadecane drops are similar.

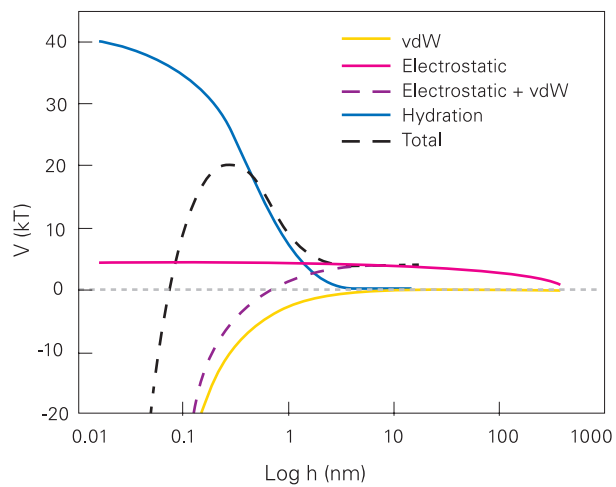

Figure 1: Interaction potentials between the drops employed in the simulations. Calculations labeled as $C_n$ (E) use van der Waals and electostatic interactions only. $C_n$ (E,G) calculations use the same interaction potentials as $C_n$ (E) but additionally include the buoyancy force (not shown). $C_n$ (E,H) simulations include van der Waals, electrostatic and hydration forces. This particular figure corresponds to dodecane drops suspended in water.

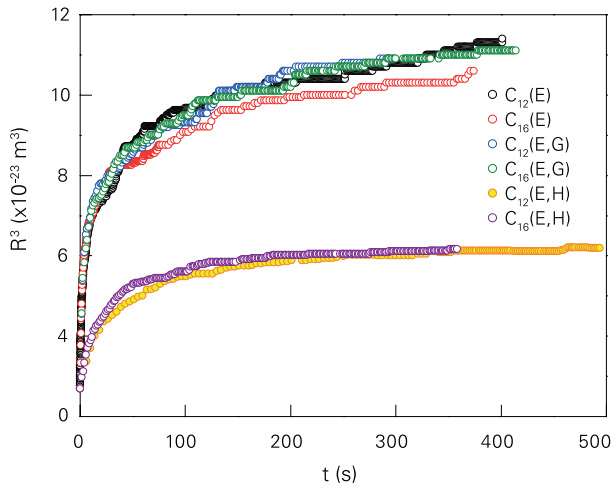

Figure 2: Change of R³ vs. t corresponding to the systems of dodecane and hexadecane.

## 4. RESULTS AND DISCUSSION

### 4.1 Change of the cube average radius as a function of time

In the following analysis, the temporal variation of the cube average radius found in the simulations will be referred as $V_{obs}$:

$$V_{obs} = dR_a^3/dt \qquad (18)$$

Figure 2 shows the variation of $R_a^3$ as a function of time for the set of systems containing either dodecane or hexadecane. As expected, the main contribution to the observed behavior during the first 400 seconds is due to the flocculation and the coalescence of the drops. The curves are not completely smooth and instead the data oscillate around an average slope. As we demonstrated before (Urbina-Villalba, 2009a, 2009c, 2012) this occurs because the exchange of molecules between the drops decreases the average radius of the dispersion, while the elimination of the drops increases it. Hence, each curve of $R_a^3$ vs. t has intervals in which one process or the other predominates. Table 2 shows the asymptotic values of $V_{obs}$ obtained in these calculations. All asymptotic values of $V_{obs}$ are positive. In fact, the asymptotic values of $V_{obs}(C_{12})$ and $V_{obs}(C_{16})$ are of the same order of magnitude, and very similar for equivalent systems.

The values of $V_{obs}$ reported by Sakai et al. [2002] for octane, dodecane and hexadecane were $1.9 \times 10^{-21}$ m³/s, $4.0 \times 10^{-26}$ m³/s, $1.2 \times 10^{-27}$ m³/s, respectively. The theoretical LSW reported by Sakai et al. for these systems are: $2.8 \times 10^{-24}$ m³/s, $1.3 \times 10^{-26}$ m³/s, $7.0 \times 10^{-29}$ m³/s, respectively. For this purpose the authors used data of the mean average radius corresponding to 45, 110, and 5000 *minutes* (see Figs. 2 and 3 in [Sakai, 2002]). Notice that the largest time used in the computation of the values of $V_{obs}$ reported in Table 2 is only 686 seconds. The highest discrepancy is observed for the octane-water systems. The simulations suggest rates of the order of $10^{-24}$ m³/s similar to the LSW prediction. However the value measured by Sakai et al. is three orders of magnitude larger.

It is important to remark at this point that in order to sample interaction potentials that are in the order of a few nanometers, a very small time step should be used. As a consequence, our simulations are very time consuming. In fact the present results correspond to more than two years of continuous simulations.

Table 2 also shows that the values of $V_{obs}$ can also be negative during certain periods of time. These values of $V_{obs}$ can be conveniently referred as $V_{exc}$ since they are solely caused by the exchange of molecules between the drops:

$$V_{obs} < 0 \quad \Rightarrow \quad V_{obs} = V_{exc} \qquad (19)$$

Table 2: Asymptotic values of $V_{obs}$

| System | $V_{obs}$ (m³/s) | r² | Time Interval (s) | $V_{obs}$ (m³/s) | r² | Time Interval (s) |
|---|---|---|---|---|---|---|
| $C_8$ (E) | $2.1 \times 10^{-24}$ | 0.9319 | 0 < t < 442 | $-1.8 \times 10^{-24}$ | 0.9877 | 407 < t < 442 |
| $C_{12}$ (E) | $5.8 \times 10^{-26}$ | 0.9898 | 260 < t < 401 | $-1.6 \times 10^{-27}$ | 1.0000 | 390 < t < 398 |
| $C_{16}$ (E) | $3.4 \times 10^{-26}$ | 0.9525 | 139 < t < 376 | $-7.2 \times 10^{-30}$ | 1.0156 | 373 < t < 376 |
| $C_8$ (E,G) | $2.0 \times 10^{-24}$ | 0.9570 | 0 < t < 665 | $-1.8 \times 10^{-24}$ | 0.9999 | 654 < t < 651 |
| $C_{12}$ (E,G) | $2.4 \times 10^{-26}$ | 0.9853 | 195 < t < 316 | $-1.3 \times 10^{-27}$ | 1.0000 | 284 < t < 295 |
| $C_{16}$ (E,G) | $3.0 \times 10^{-26}$ | 0.9830 | 208 < t < 442 | $-9.4 \times 10^{-30}$ | 1.0086 | 409 < t < 414 |
| $C_8$ (E,H) | $2.7 \times 10^{-24}$ | 0.9449 | 0 < t < 541 | $-2.3 \times 10^{-24}$ | 0.9884 | 513 < t < 541 |
| $C_{12}$ (E,H) | $1.1 \times 10^{-26}$ | 0.9598 | 197 < t < 686 | $-1.4 \times 10^{-27}$ | 1.0000 | 546 < t < 594 |
| $C_{16}$ (E,H) | $1.0 \times 10^{-26}$ | 0.9240 | 238 < t < 358 | $-6.1 \times 10^{-30}$ | 0.9989 | 352 < t < 358 |

As previously remarked, the increase of the average radius is only caused by the coalescence and/or the dissolution (elimination) of the drops. However, the exchange of molecules between drops also occurs during the increase of the radius but at a much lower rate. Hence the positive values of $V_{obs}$ cannot be solely attributed to the coalescence and elimination of the drops:

$$V_{obs} > 0 \quad \Rightarrow \quad V_{obs} \approx V_{CD} \qquad (20)$$

Where subscripts C and D stand for Coalescence and Dissolution, respectively. However, if the process of ripening is suppressed:

$$V_{obs} > 0 \quad \text{and} \quad V_{OR} = 0 \quad \Rightarrow \quad V_{obs} = V_C \qquad (21)$$

This can be achieved using a very insoluble of oil, or mixing the alkane with a very insoluble substance (like squalene) before preparing the emulsion.

As Table 2 shows, the rate of decrease of the cube average radius during the exchange of molecules ($V_{exc}$) is very different for the three oils, changing in three orders of magnitude between octane and dodecane, and also between dodecane and hexadecane. This is due to the differences in the solubility of the oils which decreases in two order of magnitude as the number of carbon atoms of the hydrophobic chain is increased by four (Table 1).

The above findings make us wonder if it is possible to reach a stable average radius as a combination of the referred opposing trends. This does not appear feasible for the case of octane where the positive and negative values of $V_{obs}$ are of the same order of magnitude. However, a metastable situation could be realistically expected in the case of dodecane and hexadecane, especially because both the flocculation rate and the Ostwald ripening rate decrease with the augment of the particle radius. Nevertheless, as the drops coalesce the number of particles per unit volume diminishes, additionally decreasing the flocculation rate. Therefore, a careful observation of the temporal behavior of the DSD of these systems indicates that Ostwald ripening predominates at longer times (see below).

Since the differences between the Hamaker constants of the three oils are small, and the surface charge of the drops is assumed to be the same (Table 1), the interaction potentials between the drops are similar for equivalent systems (Figure 1). Hence, the values of $V_{obs}$ for $C_{12}$ (E) and $C_{16}$ (E) are comparable. Moreover, the introduction of the gravity force does not change $V_{obs}$ considerably, because the drops are too small to be substantially displaced by gravity. Therefore $V_{obs}$ ($C_{12}$ (E)) ~ $V_{obs}$ ($C_{12}$ (E,G)) ~ $V_{obs}$ ($C_{16}$ (E)) ~ $V_{obs}$ ($C_{16}$ (E,G)). The situation is different for drops of several hundred nanometers, especially if the drops form stable aggregates.

If a short-range potential (like a hydration potential) is introduced, the coalescence of the drops is delayed, therefore $R_a^3$ increases at a considerable slower rate. That is why the curves of $R_a^3$ vs. t for $C_{12}$ (E,H) and $C_{16}$ (E, H) lay significantly lower than the ones corresponding to the rest of the simulations. It is important to realize that by definition; the calculation of $V_{obs}$ presented in this work is different from the value of $dR_a^3/dt$ experimentally computed. The simulations keep track of the size variation of all drops. Hence, Eq. (18) can be evaluated unambiguously. Instead, most experimental instruments (if not all) measure the average size of the *aggregates* of the dispersion. Either because they assign a hydrodynamic radius to anything that scatters light, or equivalently, because they determine the diffusion coefficient of the aggregates present in the emulsion and assign a hydrodynamic radius to them. Since the repulsive potentials between drops promote a considerable degree of aggregation, the experimental measurements evaluate $dR_a^3/dt$ using the size of the aggregates instead of the size of the drops:

$$R_{a,exp} = \frac{1}{N_{agg}} \sum_1^{agg} R_1 \qquad (22)$$

Curiously, the asymptotic values of $V_{obs}$ ($C_{12}$ (E,H)) and $V_{obs}$ ($C_{16}$ (E, H)) are of the same order of magnitude than the ones of $V_{obs}$ ($C_{12}$ (E)) and $V_{obs}$ ($C_{16}$ (E)), respectively (see Table 2).

Figure 3 shows the results of $R_a^3$ vs. t for the octane systems. These results are completely different from the ones of dodecane and hexadecane. The cubic average radius increase much more pronouncedly but in a non-monotonic way. The average slope of the curve is roughly linear but the curve shows a saw-tooth variation.

In the case of dodecane and hexadecane, the coalescence of the drops prevents the elimination of drops by complete dissolution. The rate of coalescence increases the average radius faster than the decrease caused by molecular exchange. This can be confirmed because the program writes a separate file with the information of the drops which that are eliminated by the process of Ostwald ripening. This file is not even created in the case of dodecane and hexadecane, but it contains a large amount of drops in the case of octane.

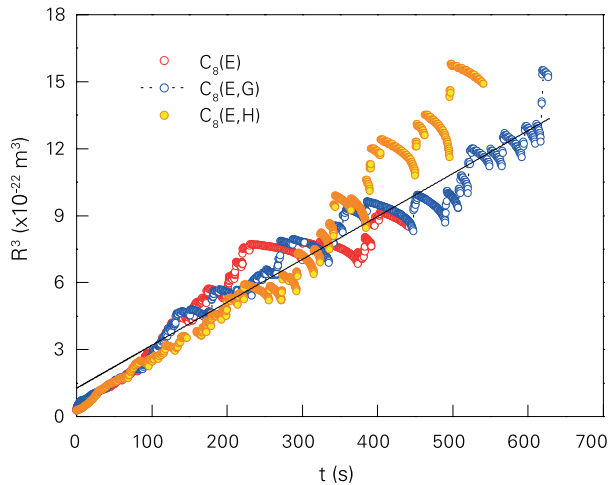

Figura 3: Change of $R^3$ vs. t exhibited by octane-in-water nanoemulisions.

TABLE 3: Mixed Flocculation-Coalescence Rates ($k_{FC}$)

| System | $k_{FC}$ (m³/s) | time | $r^2$ |
|---|---|---|---|
| $C_8$ (E) | 8.3 × 10⁻²⁰ | (0 s < t < 15 s) | 0.8403 |
|  | 5.6 × 10⁻²⁰ | (23 s < t < 95 s) | 0.9771 |
|  | 2.8 × 10⁻²⁰ | (120 s < t < 442 s) | 0.8859 |
| $C_{12}$ (E) | 5.2 × 10⁻²¹ | (0 s < t < 252 s) | 0.8501 |
|  | 1.0 × 10⁻²¹ | (252 s < t < 401 s) | 0.9918 |
| $C_{16}$ (E) | 3.5 × 10⁻²¹ | (0 s < t < 368 s) | 0.8027 |
| $C_8$ (E,G) | 6.9 × 10⁻²⁰ | (0 s < t < 22 s) | 0.8747 |
|  | 5.6 × 10⁻²⁰ | (22 s < t < 122 s) | 0.9787 |
|  | 4.4 × 10⁻²⁰ | (122 s < t < 617 s) | 0.9933 |
| $C_{12}$ (E,G) | 6.7 × 10⁻²¹ | (0 s < t < 192 s) | 0.8139 |
|  | 4.7 × 10⁻²² | (193 s < t < 316 s) | 0.9838 |
| $C_{16}$ (E,G) | 4.8 × 10⁻²¹ | (0 s < t < 215 s) | 0.6448 |
|  | 5.4 × 10⁻²² | (215 s < t < 442 s) | 0.9883 |
| $C_8$ (E,H) | 5.6 × 10⁻²⁰ | (0 s < t < 25 s) | 0.9799 |
|  | 4.3 × 10⁻²⁰ | (25 s < t < 147 s) | 0.9949 |
|  | 6.6 × 10⁻²⁰ | (160 s < t < 541 s) | 0.9848 |
| $C_{12}$ (E,H) | 5.6 × 10⁻²² | 0 s - 686 s | 0.6021 |
| $C_{16}$ (E,H) | 1.4 × 10⁻²¹ | (0 s < t < 358 s) | 0.7221 |
| $C_{12}$ (E, 0.5M) | 6.8 × 10⁻¹⁸ | (0 s < t < 0.22 s) | 0.9987 |
|  | 6.9 × 10⁻¹⁸ | (0.22s < t < 1.07 s) | 0.9989 |
|  | 7.5 × 10⁻¹⁸ | (1.07 s < t < 211 s) | 0.9977 |

Out of all disappearing drops either by coalescence or dissolution, the percentage of drops of octane which is eliminated by dissolution progressively increases with time. It is less than 10% during the first 20 seconds of calculation, 40% during the following 100 seconds, and reaches 92% after 300 seconds.

### 4.2 $k_{FCO}$ rates

Table 3 shows the results of fitting the temporal variation of the number of particles per unit volume to the equation of Smoluchowski [1917]:

$$n = \frac{n_0}{1 + k_{FCO}\, n_0\, t} \quad (23)$$

Where $k_{FCO}$ stands for a mixed flocculation rate which contains the effects of flocculation, coalescence and Ostwald ripening. In a recent paper [Urbina-Villalba, 2012] we showed that in the absence of flocculation and coalescence the rate constant of Eq. (23) also measures the rate of Ostwald ripening, where:

$$k_{FCO} \rightarrow k_O = \frac{16\,\pi\,\alpha\,D_m\,C_\infty}{27\,\phi} \quad (24)$$

Here $\alpha$, $D_m$, $C_\infty$, and $\phi$ stand for the capillary length, the diffusion constant of the oil molecules, the solubility of the oil and the volume fraction of internal phase of the dispersion.

Thus, it is possible to fit Eq. (23) to the simulation data resulting from a combination of the processes of flocculation and coalescence [Urbina-Villalba, 2000, 2004a, 2004b, 2009b; Osorio, 2011; Rahn-Chique, 2012a, 2012b, 2012c], Ostwald ripening [Urbina-Villalba, 2012] or to a combination of the three phenomena. Since the calculations with Oswald ripening include the regeneration of the DSD when $N(t') = N_0/4$, Eq. (23) can only be adjusted to the data obtained between $N = 500$ and $N = 125$. That is why Table 3 sometimes contains several set of constants for each particular system. Faster ripening rates lead to a rapid decrease of the number of particles, and more regenerations of the DSD, and vice versa.

As the regression coefficients of Table 3 show, the equation of Smoluchowski can hardly be applied to these systems. This was expected since the drops of the systems studied exhibit a considerable repulsive force. However, sometimes the regression coefficients improve considerably with time, as a result of the decrease of: a) the number of particles per unit volume, and b) the diffusion constants of the remaining drops. According to Table 3 the values of $k_{FCO}$ roughly fluctuate between $10^{-20}$ and $10^{-22}$ m³/s for the present systems.

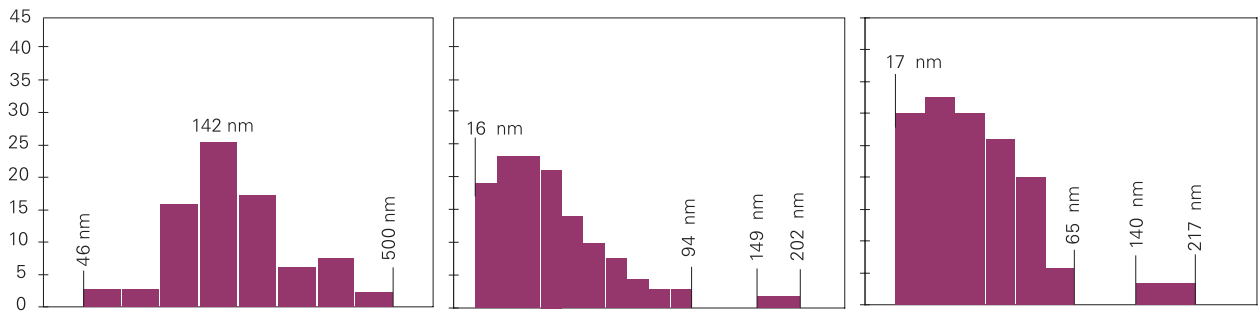

Figure 4: Qualitative form of the distributions obtained by Sakai *et al*. [2002] for octane, dodecane and hexadecane in-water nanoemulsions at t = 4 minutes.

TABLE 4: Parameters resulting from the fitting of Eq. (25) to the simulation data.

| Sistema | A | B | $k_F$ (m³/s) | $k_C$ (m³/s) |
|---|---|---|---|---|
| $C_8$ (E) | 0.71<br>0.73<br>0.62 | 0.29<br>0.27<br>0.38 | $6.2 \times 10^{-19}$<br>$3.6 \times 10^{-20}$<br>$1.2 \times 10^{-19}$ | $1.5 \times 10^{-21}$<br>$5.0 \times 10^{-20}$<br>$4.0 \times 10^{-21}$ |
| $C_{12}$ (E) | 0.7 | 0.3 | $5.0 \times 10^{-19}$ | $5.0 \times 10^{-22}$ |
| $C_{16}$ (E) | 0.7 | 0.3 | $5.5 \times 10^{-19}$ | $2.5 \times 10^{-22}$ |
| $C_8$ (E, G) | 0.7<br>0.7<br>0.59 | 0.3<br>0.3<br>0.41 | $5.0 \times 10^{-19}$<br>$3.0 \times 10^{-20}$<br>$7.1 \times 10^{-20}$ | $6.0 \times 10^{-21}$<br>$6.0 \times 10^{-20}$<br>$1.5 \times 10^{-20}$ |
| $C_{12}$ (E,G) | 0.7<br>0.55 | 0.3<br>0.45 | $5.0 \times 10^{-19}$<br>$8.0 \times 10^{-22}$ | $5.0 \times 10^{-22}$<br>$1.0 \times 10^{-22}$ |
| $C_{16}$ (E,G) | 0.72<br>0.70 | 0.28<br>0.30 | $6.5 \times 10^{-19}$<br>$8.0 \times 10^{-22}$ | $3.5 \times 10^{-23}$<br>$2.0 \times 10^{-22}$ |
| $C_8$ (E, H) | 0.50<br>0.69<br>0.64 | 0.50<br>0.31<br>0.36 | $2.5 \times 10^{-20}$<br>$7.6 \times 10^{-20}$<br>$8.0 \times 10^{-20}$ | $4.2 \times 10^{-20}$<br>$1.2 \times 10^{-20}$<br>$2.4 \times 10^{-20}$ |
| $C_{12}$ (E, H) | 0.57 | 0.43 | $4.6 \times 10^{-20}$ | $5.0 \times 10^{-23}$ |
| $C_{16}$ (E,H) | 0.54 | 0.46 | $1.0 \times 10^{-19}$ | $1.3 \times 10^{-22}$ |

Table 4 shows the set of "reaction" rates resulting from the fitting of Eq. (25) to the simulation data [Urbina-Villalba, 2005]:

$$n(t) = n_0 \left[ A/(1 + k_1 n_0 t) + B \exp(-k_2 n_0 t) \right] \quad (25)$$

Here A, B, $k_1$ and $k_2$ are constants and B = 1 - A. The above equation was formerly deduced to describe the variation of the total number of aggregates in an emulsion in the presence of a significant repulsive potential. Constants $k_1$ and $k_2$ were formerly ascribed to the rates of flocculation and coalescence, which are represented by the first and the second terms on the right hand side of Eq. (25), respectively. However, it was soon realized that the exponential term was also able to fit the terminal flocculation which occurs after the process of coalescence stops. Moreover, two exponential terms can be used instead of one.

This is important because Eq. (25) may serve to evaluate the rates of the destabilization process in the cases were Eq. (23) does not follow the experimental trend. Unfortunately it is not possible to predict the number of drops of each size consistent with Eq. (25), while in the case of Eq. (23) an analytic expression for the number of aggregates of each size is available [Smoluchowski, 1917].

The calculations that justified the use of Eq. (25) considered the redistribution of surfactant molecules among the surviving drops [Urbina-Villalba, 2005]. This causes an increase of the surface charge with the size of the drops which favours the stabilization of the system. In the present case, the surface charge of the drop is constant, but the total charge still increases with the coalescence of the drops. Moreover, the diffusion constant of the drops diminishes with size. As a consequence, the curves of $R^3$ *vs*. t develop an asymptotic stabilization, similar to the one formerly found (Figure 2). Consequently Eq. (25) fits the data on the number of drops. Remarkably in this case the process of Ostwald ripening is also taken into account, and even in the case of octane a good agreement with the calculations is found.

4.2 Drop Size Distributions

Figure 4 shows the qualitative forms of the first drop size distributions measured by Sakai *et al*. [2002] for octane, dodecane and hexadecane oil-in-water nanoemulsions. The first two DSD correspond to t = 4 minutes after their preparation, and the one of hexadecane, to t = 10 minutes. It is observed that the form of the DSD corresponding to octane is similar to a wedding cake and it is very different from the rest. This distribution is very broad, spanning from 46 and 500 nm. It is roughly symmetric with a principal peak around 142 nm. Instead, the DSD of dodecane and hexadecane consist in two parts. The principal signal has the shape of a log normal distribution with a right-skewed skirt. The main signal of dodecane is located between 16 and 94 nm and the one of hexadecane between 17 and 65 nm. Their zeniths are located around 23 and 26 nm, respectively. A secondary contribution of much lower frequency is located between 149 to 202 nm for dodecane and between 140 and 217 nm in the case of hexadecane.

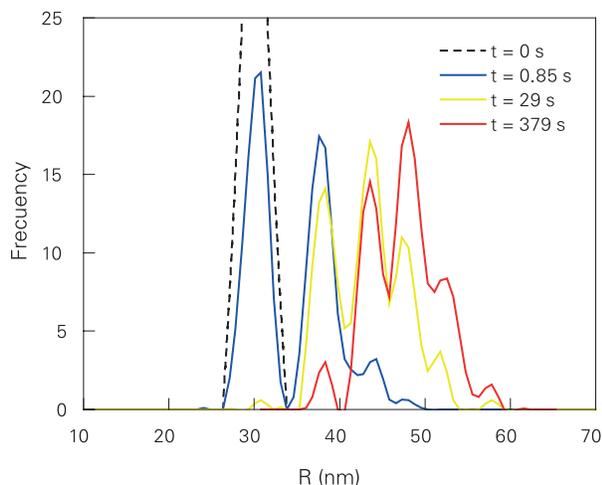

Figure 5: Initial evolution of the DSD corresponding to the $C_{16}$ (E) system.

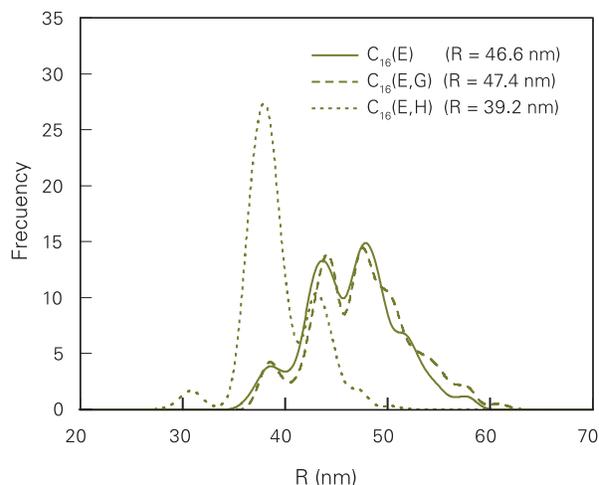

Figure 6: DSD of $C_{16}$ (E), $C_{16}$ (E, G) and $C_{16}$ (E, H) systems at t = 4 minutes.

The algorithm of Ostwald ripening implemented in the ESS program (De Smet *et al.*, 1997) reproduces the predictions of the LSW theory in the stationary regime. In a previous paper we showed simulations in which the drops were fixed in the space, and only the process of Ostwald ripening proceeded [Urbina-Villalba, 2009a]. In that case a rather long time step was used ($\Delta t = 5.4 \times 10^{-5}$ s, M = 0.05). We repeated those simulations using $\Delta t = 2.2 \times 10^{-8}$ s (M = 2.04 x $10^{-5}$) obtaining similar results (see Figure 4 of Ref. [Urbina-Villalba, 2009a]. A dodecane-in-water emulsion which starts from a Gaussian distribution centred at $R_a$ = 30 nm, reaches a value of 1.1 x $10^{-26}$ m$^3$/s ($r^2$ = 0.9932) after 1700 s At this time, the DSD is skewed to the left, showing a long tail that spans from 42.8 nm (maximum value) to 6.8 nm, with a mean particle size of $R_a$ = 31.2 nm. During the transient period the value of $R_a$ decreases first, passes through a minimum, and only increases after 987 s, when the first drop of oil is lost by dissolution.

When the drops are able to flocculate and coalesce, the DSD of a dodecane-in-water develops a right skirt in a fraction of a second [Urbina-Villalba, 2009a, b, c; Urbina-Villalba 2012]. This behavior is illustrated in Figure 5 for a dilute ($\phi$= 2 x $10^{-4}$) hexadecane-in-water emulsion (system $C_{16}$ (E)). In the absence of surfactants, flocculation and coalescence occur so fast that the elimination of small drops due to molecular exchange does not occur during the first four minutes. As the number of particles per unit volume (n) and the diffusion constant of the remaining particles decreases ($D_{eff,i}$), the curve of $R_a^3$ vs. t decreases its slope approximating a horizontal line (see $C_{12}$ (E) in Figure 2). This occurs because the exchange of molecules of oil between drops starts to balance the increase of the average size due to the coalescence of drops. As a result, the DSD exhibit a pseudo-stationary state after 182 s. This distribution barely changes during several minutes, although it steadily progresses from right-skewed to left-skewed (see Figure 5). Notice that the form of this pseudo-stationary distribution is different from the one predicted by LSW. In fact, the DSD corresponding to different times coincide without the need of scaling each particle size.

Figures 6, 7 and 8 show the DSD corresponding to all systems simulated for t = 240 s. In all three figures the DSDs of $C_n$ (E) systems and the ones of $C_n$ (E, G) simulations are similar in position and form. This confirms that for $R_i$ < 70 nm the effect of gravity is not pronounced. This was previously observed in the $R^3$ vs. t plots. Instead, the short-range hydration potential causes a substantial decrease in the movement of the DSD towards larger radii. It is for this reason that the principal peak of the $C_n$ (E,H)

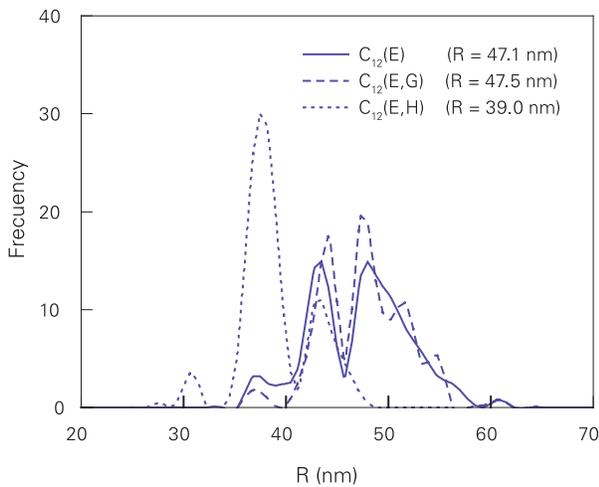

Figure 7: DSD of $C_{12}$ (E), $C_{12}$ (E, G) and $C_{12}$ (E, H) systems at t = 4 minutes.

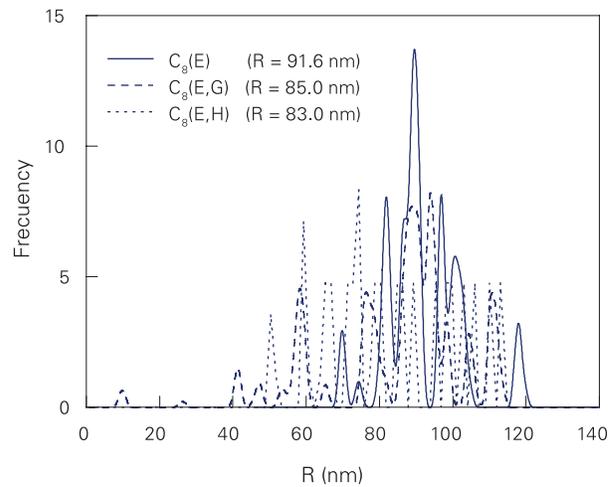

Figure 8: DSD of $C_8$ (E), $C_8$ (E, G) and $C_8$ (E, H) systems at t = 4 minutes.

systems lay below $R_i$ = 40 nm at t = 240 s, which makes it distinctively separated from the rest of the distributions. It is also clear that the distributions corresponding to octane are wider than the rest, and totally polydisperse. The qualitative form of the distribution is similar to the one reported by Sakai et al. [2002] for this system, but its average size is not correct.

Figure 9 shows a comparison between all $C_n$ (E) at a time of four minutes. It is observed that the DSDs of the systems of $C_{12}$ (E) and $C_{16}$ (E) are similar and lay at the same position in particle size. Instead, the distribution of octane is clearly located at larger particle size. The comparison between the systems of octane, dodecane and hexadecane corresponding to $C_n$ (E, G) and $C_n$ (E, H) show a similar trend.

## 5. CONCLUSIONS

In its way toward phase separation, an emulsion undergoes several processes of destabilization including flocculation, coalescence, creaming and Ostwald ripening. These processes occur simultaneously and influence each other. However, the behaviour of the emulsion depends markedly on the size of its drops, the solubility of its oil, and the interaction potential between the drops. The size of the drops determines if the dispersion will remain homogeneous as a function of height or will cream. It also establishes if the drops will deform upon flocculation. The solubility of the oil sets up the Oswald ripening rate. And last but not least, the interaction potential determines the rate of aggregation, and consequently, the rate of coalescence.

In the oil-in-water emulsions studied the size of the drops is too small to be affected by gravity. Consequently, the behaviour of the dispersion depends on a competition between Ostwald ripening and the mixed process of flocculation and coalescence. According to our simulations and at least during the first 240 seconds, the effect of flocculation and coalescence is decisive. The average radius of the emulsion only increases when the number of drops diminishes either by coalescence or dissolution. The molecular exchange between drops produced by ripening only causes a decrease of the average radius unless a drop is eliminated. As a consequence, the curves of $R^3$ vs. t -often used to characterize the Oswald ripening rate-, appear to oscillate around an average slope.

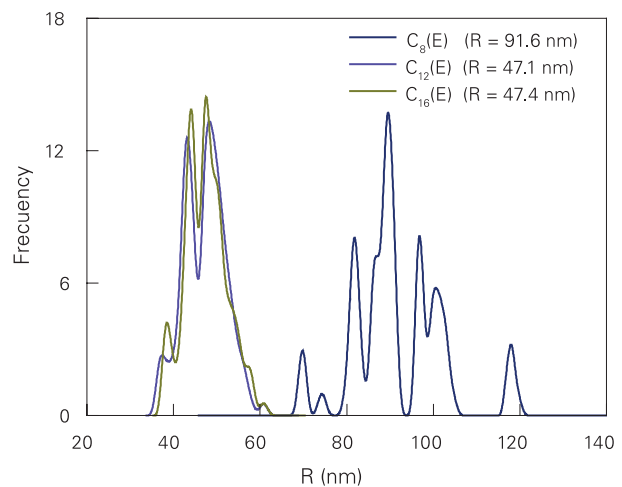

Figure 9: DSD of $C_8$ (E), $C_{12}$ (E) and $C_{16}$ (E) systems at t = 4 minutes.

In the case of octane, its high solubility causes a pronounced degree of molecular exchange. Drops are lost by dissolution and coalescence. The rates of $V_{obs} > 0$ and $V_{obs} < 0$ are similar and of the order of $10^{-24}$ m$^3$/s. However, even in this case, the variation of $R^3$ vs. t shows a concave downwards curve. This curve evidences that as the drops grow larger and the process of coalescence proceeds, the rate of flocculation diminishes. This is caused by the lower diffusion rate of bigger drops, and mostly by the fact that the number of drops per unit volume diminishes if the size of the drops increases at constant volume fraction.

The cases of dodecane and hexadecane are very different from octane. In these systems the rate of flocculation and coalescence is substantially larger than the one of ripening. Thus, the drops are not lost by dissolution during the first 240 seconds. Coalescence increases the average radius at a much higher rate that it decreases due to molecular exchange. Since the interaction potential between the drops of both systems is similar, the position and form of their drop size distribution coincide at several times. However, it is necessary to remark that the same surface charge of the drops was used for both systems. A different surface charge will also generate differences between the DSD of these systems. In any event, as time passes the flocculation rate diminishes due to the increase of the size of the drops and the decrease of their number density. As a result, the process of molecular exchange starts to equilibrate the effect of flocculation and coalescence. The DSD progressively change from right-skewed to left-skewed, experiencing a pseudo-stationary period in which the DSD does not change during several minutes. It is expected that the slope of $R^3$ vs. t reflect solely reflect the ripening rate at very long times, as predicted by the LSW theory for the stationary regime.

Finally, the inclusion of a hydration potential greatly diminishes the flocculation rate, favouring the earlier equilibrium between coalescence and molecular exchange. The effect of the potential is significant even though it only acts below 4 nanometers.

### 6. ACKNOWLEDGEMENTS